\documentclass{article}

\usepackage{arxiv}

\usepackage[utf8]{inputenc} 
\usepackage[T1]{fontenc}    
\usepackage{hyperref}       
\usepackage{url}            
\usepackage{booktabs}       
\usepackage{amsfonts}       
\usepackage{nicefrac}       
\usepackage{microtype}      
\usepackage{lipsum}		
\usepackage{graphicx}
\usepackage{natbib}

\usepackage{amsmath} 
\usepackage{float}
\usepackage{subcaption}
\usepackage{threeparttablex}
\usepackage{adjustbox}

\usepackage{booktabs}      
\usepackage{siunitx}       
\usepackage{threeparttable} 
\usepackage{caption}       
\usepackage{geometry}      

\usepackage{amsmath}
\DeclareMathOperator*{\argmax}{arg\,max}

\title{A multi-factor model for improved commodity pricing: Calibration and an application to the oil market}

\author{ Luca Vincenzo Ballestra\\
	Department of Statistical Sciences \\
	University of Bologna \\ Bologna, Italy \\
       \And
       Christian Tezza\thanks{
   Corresponding author. Dipartimento di Scienze Statistiche, Alma Mater Studiorum Università di Bologna, Via Belle Arti 41, 40126 Bologna, Italy, e-mail: \href{mailto:christian.tezza@unibo.it}{christian.tezza@unibo.it}.}\\
  Department of Statistical Sciences \\
	University of Bologna \\ Bologna, Italy \\
}

\renewcommand{\shorttitle}{\textit{arXiv} Template}

\hypersetup{
pdftitle={A template for the arxiv style},
pdfsubject={q-bio.NC, q-bio.QM},
pdfauthor={David S.~Hippocampus, Elias D.~Striatum},
pdfkeywords={First keyword, Second keyword, More},
}

\begin{document}
\maketitle

\begin{abstract}

We present a new model for commodity pricing that enhances accuracy by integrating four distinct risk factors:  spot price, stochastic volatility, convenience yield, and stochastic interest rates. While the influence of these four variables on commodity futures prices is well recognized, their combined effect has not been addressed in the existing literature. We fill this gap by proposing a model that effectively captures key stylized facts including a dynamic correlation structure and time-varying risk premiums. Using a Kalman filter-based framework, we achieve simultaneous estimation of parameters while filtering state variables through the joint term structure of futures prices and bond yields. We perform an empirical analysis focusing on crude oil futures, where we benchmark our model against established approaches. The results demonstrate that the proposed four-factor model effectively captures the complexities of futures term structures and outperforms existing models.

\end{abstract}


\section{Introduction}

In recent years, the crude oil market emerged as the largest and most influential commodity market globally. Accurate modeling of the key drivers of commodity price dynamics is essential for effective pricing, hedging, and risk management of commodity derivatives.



Empirically, as documented in \cite{litzenberger95}, \cite{duffie99}, and \cite{eydeland03}, many commodity prices are mean-reverting, strongly heteroscedastic, and influenced by the \cite{samuelson65} effect, whereby volatility tends to increase as futures contracts approach maturity. Furthermore, commodity futures prices are often ``backwardated,'' meaning they decline as the delivery date approaches. Notably, the analysis of crude oil prices in \cite{routledge00} shows that the degree of backwardation is positively related to volatility, implying that volatility encompasses a component spanned by futures contracts.

To address there stylized facts, affine factor models gained traction among practitioners due to their desirable analytical properties and because they allow for straightforward pricing of futures and bonds. However, a significant challenge remains in determining the optimal number and type of factors driving the spot price dynamics. In the seminal work of \cite{schwartz97}, it is assumed that all the uncertainty is summarized by one factor, namely, the spot price of the commodity. However, one-factor models imply that the returns of all futures in the term structure are perfectly correlated and that the degree of backwardation is time-invariant, which are overly restrictive assumptions inconsistent with empirical data.

To address the drawbacks of using a single factor, \cite{schwartz90} and \cite{schwartz97} considered two stochastic factors, the spot price and the convenience yield. Under these models, futures prices are no longer perfectly correlated, allowing for richer term structures. However, \cite{routledge00} argued that the correlation between the spot price and the convenience yield needs to be time-varying and that two factors do not adequately capture price volatility dynamics. Consequently, several articles have proposed three-factor models, typically including either stochastic volatility (see \cite{deng00}, \cite{geman05}, \cite{hikspoors08}, \cite{lutz10}, and \cite{hughen10}) or stochastic interest rate (e.g., \cite{cortazar94}, \cite{schwartz97}, \cite{cortazar03}, \cite{casassus05}, \cite{tang12} and \cite{mellios16}).

While the above-mentioned articles suggest that effective pricing of commodity futures require at least three factors, the filtered estimates from these models are not always satisfactory. \cite{hughen10} found that the pricing performance of three-factor models is often comparable to that of two-factor models, indicating the potential need for a fourth factor. \cite{cortazar06} support this finding showing that a fourth factor is crucial for fitting the volatility term structure of crude oil futures. Four-factor model specifications proposed by \cite{yan02}, \cite{schwartz09a}, \cite{schwartz09b}, and \cite{spinler17} aim to enhance commodity futures pricing by incorporating factors like mean reversion, convenience yield, stochastic volatility, and jump clustering, albeit with different factor combinations.
 
To shed light on the appropriate number and types of factors needed to model commodity prices, we propose a novel four-factor model that incorporates the spot price, the convenience yield, the interest rate, and the volatility of the spot price, treating the variables in the drift term of the log spot price as stochastic. This combination of factors has not been considered in the literature, yet it allows for a more accurate evaluation of the impact of interest rates on the risk-neutral drift of log spot prices. In line with \cite{hughen10} and \cite{spinler17}, we utilize affine specifications for the drift and covariance terms, while maintaining a parsimonious structure, and we account for a time-varying correlation matrix and time-varying risk premia since, following \cite{duffee02}, \cite{casassus05}, and \cite{cheridito07}, we model risk premia as affine functions of the state factors. 

We develop an innovative estimation framework that incorporates bond yields, marking the first effort to jointly model the four factors that we model using panel data on futures prices and bond yields. Moreover, we adapt the discretization derived by \cite{kelly23} for square-root processes to our state-space estimation framework to ensure the positivity of the state volatility process. We conduct rigorous empirical testing, both in-sample and out-of-sample, using a comprehensive panel dataset of crude oil futures prices and bond data. The results show that our approach outperforms popular models like \cite{schwartz97}, \cite{yan02}, \cite{hughen10}, and \cite{spinler17}, particularly when bond estimation is included, which enhances short-term futures accuracy.

The contribution of this paper is twofold. First, our four-factor model improves upon existing benchmarks for pricing commodity futures while accounting for the interaction between futures and bond yields. Second, our analysis highlights the crucial role volatility and interest rates in the drift and diffusion terms of log spot prices. Notably, the incorporation of bond yields not only enhances the accuracy of interest rate estimation, but also provides a more effective representation of commodity futures prices dynamics. While the empirical analysis in this paper focuses on the crude oil market, our model and estimation framework are applicable to a broad range of commodities.

The remainder of this article is structured as follows: In Section \ref{sec_model}, we present our model and derive formulas for the valuation of commodity futures and bond yields. Section \ref{sec_benchmarks} briefly reviews the benchmark models. Section \ref{sec_method} details the data sources and the model estimation framework. Empirical results, including both in-sample and out-of-sample analyses, are presented in Section \ref{sec_empirical}. Finally, Section \ref{sec_conclusions} concludes the paper.

\section{The four-factor model} \label{sec_model}

In this section, we present our proposed model for commodity prices, which incorporates Stochastic interest Rate and Volatility, comprising a total of four factors. Thus, we refer to this model as \textit{SRV-4f}.

We assume that the commodity spot price \(S\) is driven by four independent sources of uncertainty and is influenced on the convenience yield \(\delta\), the (instantaneous) interest rate \(r\), and the variance of the log spot price \(v\). We define the state vector \({X}(t) = \begin{pmatrix} \ln S(t) & \delta(t) & r(t) & v(t) \end{pmatrix}^\top\), which we assume satisfies the following stochastic differential equation, under the risk-neutral probability measure:

\begin{equation} \label{eq_statevar}
    d {X}(t) = [A + B {X}(t)] dt + {\Sigma}^{1/2}(t,X(t)) d {Z}(t), 
\end{equation}

\begin{equation*}
    {A} =
\begin{pmatrix}
    0 \\
    k_2 \mu_2 \\
    k_3 \mu_3 \\
    k_4 \mu_4
\end{pmatrix}, \quad 
{B} = 
\begin{pmatrix}
    0 & -1 & +1 & -\frac{1}{2}\\
    0 & -k_2 & 0 & 0 \\
    0 & 0 & -k_3 & 0 \\
    0 & 0 & 0 & -k_4 \\
\end{pmatrix},
\end{equation*}

\noindent where $dZ(t)$ is a vector of four uncorrelated standard Brownian motion increments, $\Sigma^{1/2}(t,X(t))$ denotes the Cholesky decomposition of the instantaneous variance and covariance matrix given by

\begin{equation} \label{eq_diffusion}
    {\Sigma}(t,X(t)) = {\Omega}_0 +  {\Omega}_1 v(t),
\end{equation}

\noindent where

\begin{equation*}
    {\Omega}_0 = \begin{pmatrix}
        0 & s_{12} & s_{13} & 0 \\
    s_{12} & s_{22} & s_{23} & 0  \\
    s_{13} & s_{23} & s_{33} & 0  \\
    0 & 0 & 0 & 0
    \end{pmatrix},\quad 
    {\Omega}_1  =  \begin{pmatrix}
    1 & \rho_{12} \sigma_{22} & \rho_{13} \sigma_{33} & \rho_{14} \sigma_{44} \\
    \rho_{12} \sigma_{22} & \sigma^2_{22} &\rho_{23} \sigma_{22} \sigma_{33} & \rho_{24} \sigma_{22} \sigma_{44} \\
    \rho_{13} \sigma_{33} & \rho_{23} \sigma_{22} \sigma_{33} & \sigma^2_{33} & \rho_{34} \sigma_{33} \sigma_{44} \\
    \rho_{14} \sigma_{44}& \rho_{24} \sigma_{22} \sigma_{44} & \rho_{34} \sigma_{33} \sigma_{44} & \sigma^2_{44}  \\
\end{pmatrix}.
\end{equation*}

As in \cite{duffie96}, both the drift term and \(\Sigma(t, {X}(t))\) are affine functions of the state variables. We note that \(\Omega_0(1,1)=0\) and \(\Omega_1(1,1)=1\), so that \(v(t)\) is actually the variance of the first state variable \(\log S(t)\), and we require \(\Omega_0 + \Omega_1 v(t)\) to be positive definite.

We define the vector of risk-premia \(\Lambda (t,X(t)) = \begin{pmatrix}
    \lambda_x(t,S(t)) & \lambda_\delta(t,\delta(t)) & \lambda_r(t,r(t)) & \lambda_v(t,v(t))
\end{pmatrix}^\top \) as an affine function of ${X}(t)$, that is

\begin{equation}
    {\Lambda} (t,X(t)) = {\Sigma}(t,X(t))^{-1} \left( {\widehat{A}} - {A} + [{\widehat{B}} - {B}] X(t) \right), 
\end{equation}

\noindent where the physical drift parameters ${\widehat{A}}$ and ${\widehat{B}}$ are given by

\begin{equation}
   {\widehat{A}} =
\begin{pmatrix}
    \widehat{\mu}_1\\
    \widehat{k}_2 \widehat{\mu}_2 \\
    \widehat{k}_3 \widehat{\mu}_3 \\
    \widehat{k}_4 \widehat{\mu}_4
\end{pmatrix}, \quad 
{\widehat{B}} =
\begin{pmatrix}
    0 & -1 & 0 & - \frac{1}{2}\\
    0 & -\widehat{k}_2 & 0 & 0 \\
    0 & 0 & -\widehat{k}_3 & 0 \\
    0 & 0 & 0 & -\widehat{k}_4 \\
\end{pmatrix}.
\end{equation}

From the Girsanov Theorem, see Theorem 11.3 in \cite{bjork09}, we know that the process $Z(t)^\mathbb{P}$, defined by

\begin{equation} \label{eq_girsanov}
    Z(t)^\mathbb{P} = Z(t) - \int_0^t \Lambda (s,X(s)) ds,
\end{equation}

\noindent is a standard $\mathbb{P}$-Wiener process, where $\mathbb{P}$ denotes the physical probability measure.

Substituting \eqref{eq_girsanov} into \eqref{eq_statevar} yields the continuous time model dynamics under $\mathbb{P}$:

\begin{equation} \label{eq_statevar_p}
    d {X}(t) = \left[\widehat{A} + \widehat{B} {X}(t) \right] dt + {\Sigma}^{1/2}(t,X(t)) d {Z}^{\mathbb{P}}(t).
\end{equation}

From a mathematical standpoint, the diffusion term in \eqref{eq_diffusion} follows an affine specification similar to the model proposed by \cite{spinler17}. However, we model different risk factors and we allow for a more parsimonious parametrization of the expressions for \(B\) and \(\Omega_0\) (see Section \ref{sec_shone}).

\subsection{Futures prices}


The futures price at time $t$ with time to maturity $\tau=T-t$ can be computed as the risk-neutral expectation of the futures price:

\begin{equation*}
    F(t, \tau, X(t)) = \mathbb{E} \left[ e^{X(t+\tau)} \middle| \mathcal{F}(t) \right], 
\end{equation*}

\noindent where $\mathcal{F}(t)$  denotes the information available at time $t$. The futures price is then the solution to the following partial differential equation (PDE) (see Proposition 5.5 in \cite{bjork09}):

\begin{equation} \label{eq_pde}
     \frac{\partial F(t,\tau, X(t))}{\partial \tau} =  
     \left(\nabla_x F(t,\tau, X(t)) \right)^\top [A+B X(t)]
    + \frac{1}{2} \text{Tr}\left\{  \left(H_x F(t,\tau, X(t))\right) {\Sigma}(t, X(t)) \right\}, 
\end{equation}

\noindent with the terminal condition \(F(t, 0, {X}(t)) = e^{{X}(t)}\), where \(\nabla_x F(t, \tau, {X}(t))\) and \(H_x F(t, \tau, {X}(t))\) are the gradient of \(F(t, \tau, {X}(t))\) with respect to \({X}(t)\) and the Hessian matrix of \(F(t, \tau, {X}(t))\) with respect to \({X}(t)\), respectively, and \(\text{Tr}\) is the trace operator.

Given that the drift and diffusion term in \eqref{eq_statevar} are affine functions of $X(t)$, the solution to equation \eqref{eq_pde} can be guessed as follows:

\begin{equation} \label{eq_pde_sol}
    \ln F(t, \tau,  X(t) ) = {\alpha(\tau) + {\beta}(\tau) {X}(t)},
\end{equation}

\noindent where ${\beta}(\tau) = \begin{pmatrix}
\beta_{1}(\tau) & 
\beta_{2}(\tau) &
\beta_{3}(\tau) &
\beta_{4}(\tau)
\end{pmatrix}$. By substituting equation \eqref{eq_pde_sol} into \eqref{eq_pde} we obtain

\begin{equation*}
    \frac{d \alpha(\tau)}{d\tau} +  \frac{d {\beta(\tau)}}{d\tau}{X}(t) = {\beta(\tau)} [A + B X(t)] + \frac{1}{2} \left[ \beta(\tau) \Omega_0 \beta(\tau)^\top+ \beta(\tau) \Omega_1 \beta(\tau)^\top v(t) ) \right].
\end{equation*}

Separation of variables leads to the following system of ordinary differential equations (ODEs):

\begin{align} \label{eq_system_ode1} 
\frac{d\alpha(\tau)}{d\tau} = \beta(\tau) A + \frac{1}{2} \beta(\tau) \Omega_0 \beta(\tau)^\top, \ &\frac{d\beta_1(\tau)}{d\tau} = \beta(\tau) B_{(1)}, \ \frac{d\beta_2(\tau)}{d\tau} = \beta(\tau) B_{(2)}, \ \frac{d\beta_3(\tau)}{d\tau} &= \beta(\tau) B_{(3)}, \\
&\frac{d\beta_4(\tau)}{d\tau} = \beta(\tau) B_{(4)} + \frac{1}{2} \beta(\tau) \Omega_1 \beta(\tau)^\top, \label{eq_system_ode5}
\end{align}

\noindent where $B_{(i)}$ denotes the $i$-th column of $B$, $i=1,2,3,4$. Equations \eqref{eq_system_ode1}-\eqref{eq_system_ode5} must be solved with terminal conditions $\alpha(0)=0,\beta_1(0)=1,\beta_2(0)=0,\beta_3(0)=0$ and $\beta_4(0)=0$.

\subsection{Bond yields}


The present value at time $t$ of a unit discount bond $P(t,\tau,X(t))$ with time to maturity $\tau$ can be computed as: 

\begin{equation} \label{eq_cond_pde}
    P(t,\tau,X(t)) = \mathbb{E}\left[e^{-\int_t^{t+\tau} r(s) ds} \middle| \mathcal{F}(t) \right].
\end{equation}

The conditional expectation \eqref{eq_cond_pde} is the solution of the following PDE:

\begin{align} \label{eq_bond_pde}
    P(t,\tau,X(t)) r(t)  = -\frac{\partial P(t,\tau,X)}{\partial \tau} &+ 
    \left(\nabla_x P(t,\tau, X(t)) \right)^\top [{A+B}{X}(t)]\\ \notag
    &+ \frac{1}{2} \text{Tr}\left\{  \left(H_x P(t,\tau, X(t))\right) {\Sigma}(t, X(t)) \right\},
\end{align}

\noindent with terminal condition $P(t,0, X(t)) = 1$. The PDE  \eqref{eq_bond_pde} can be solved as

\begin{equation} \label{eq_sol_bond}
    \ln P(t, \tau, X(t)) = {\gamma(\tau) + {\zeta}(\tau) {X}(t)},
\end{equation}

\noindent where ${\zeta}(\tau) = \begin{pmatrix}
\zeta_{1}(\tau) & 
\zeta_{2}(\tau) &
\zeta_{3}(\tau) &
\zeta_{4}(\tau)
\end{pmatrix}$. By substituting equation \eqref{eq_sol_bond} into \eqref{eq_bond_pde} we obtain

\begin{equation*}
    \frac{d \gamma(\tau)}{d\tau} +  \frac{d \zeta(\tau)}{d\tau} {X}(t) = \zeta(\tau) [{A} +  {B} {X}(t)] + \frac{1}{2} \left[  \zeta(\tau) \Omega_0 \zeta(\tau)^\top + \zeta(\tau) \Omega_1 \zeta(\tau)^\top v(t) \right] - r(t),
\end{equation*}

\noindent which leads to the following system of ODEs:

\begin{align*}  
\frac{d\gamma(\tau)}{d\tau} = {\zeta(\tau)} A +\frac{1}{2} {\zeta(\tau)\Omega_0\zeta(\tau)^\top}, \
&\frac{d{\zeta_1(\tau)}}{d\tau} = {\zeta(\tau)}{B}_{(1)}, \
\frac{d{\zeta_2(\tau)}}{d\tau} = {\zeta(\tau)}{B}_{(2)}, \
\frac{d{\zeta_3(\tau)}}{d\tau} = {\zeta(\tau)}{B}_{(3)} - 1, \\
&\frac{d{\zeta_4(\tau)}}{d\tau} = {\zeta(\tau)}{B}_{(4)} + \frac{1}{2} {\zeta(\tau)} {\Omega}_1 {\zeta(\tau)}^\top, 
\end{align*}

\noindent and $\gamma(0)= {\zeta}(0) = 0$ so that $P(t,0,X(t)) = 1$.  

\section{Benchmark models} \label{sec_benchmarks}

In this section, for the readers convenience, we briefly recall well-known factor models for commodity prices which we will utilize as benchmarks.

\subsection{Schwartz's (1997) one-factor model}

The one-factor model of \cite{schwartz97}, hereafter \textit{SCH-1f}, assumes that $\ln S(t)$ follows an Ornstein-Uhlenbeck stochastic process, under the risk-neutral measure:

\begin{equation*}
    d \ln S(t) = \kappa \left( \alpha - \lambda - \ln S(t)  \right)dt + \sigma dZ(t),
\end{equation*}

\noindent where $\lambda$ is the (constant) market price of risk and $dZ(t)$ denotes the  increment to a standard Brownian motion. 

According to equation (8) of \cite{schwartz97}, we can compute the logarithm of the futures price with time to maturity $\tau$ as:

\begin{equation*}
    \ln{F(t, \tau, S(t))} = e^{-\kappa \tau } \ln S(t) + (1 - e^{-\kappa \tau }) (\alpha - \lambda) + \frac{\sigma^2}{4 \kappa} (1 - e^{-2 \kappa \tau}).
\end{equation*}

\subsection{Schwartz's (1997) two-factor model}

The two-factor model of \cite{schwartz97}, denoted \textit{SCH-2f}, models the logarithm of the spot price and convenience yield $\delta$ via the following equations, under the risk-neutral measure:

\begin{align*}
    &d \ln S(t) =  \left(r - \delta(t) - \frac{1}{2} \sigma^2  \right) dt + \sigma_1 dZ_{1}(t), \\ \notag
    &d\delta(t) = \kappa [\left(\alpha - \delta(t) \right) - \lambda] dt + \sigma_2 dZ_{2}(t),
\end{align*}

\noindent where the increments of the standard Brownian motions are correlated with $ \mathbb{E}[dZ_{1}(t) dZ_{2}(t)] = \rho dt$. The logarithm of the futures price with time to maturity $\tau$ is given as in equation (19) of \cite{schwartz97}:

\begin{equation*}
    \ln{F(t,\tau, S(t), \delta(t))} =  \ln S(t) - \delta(t) \frac{1-e^{-\kappa \tau}}{ \kappa } + A^*(\tau),
\end{equation*}

\noindent where $A^*(\tau)$ is specified in equation (20) of \cite{schwartz97}.  

\subsection{Schwartz's (1997) stochastic interest rate model}

The three-factor model of \cite{schwartz97}, which we denote by \textit{SCH-3f}, comprises the following stochastic differential equations under the risk-neutral measure:

\begin{align*}
    &d \ln S(t) =  \left(r(t) - \delta(t) - \frac{1}{2} \sigma^2  \right)dt + \sigma_1 dZ_{1}(t), \\ 
    &d\delta(t) = \kappa \left( \alpha - \delta(t) \right)dt + \sigma_2 dZ_{2}(t), \\ 
    &dr(t) = a \left(\mu - r(t) \right) dt + \sigma_3 dZ_{3}(t),
\end{align*}

\noindent where the Brownian motion increments are correlated via $ \mathbb{E}[dZ_{1}(t) dZ_{2}(t)] = \rho_{1} dt$, $ \mathbb{E}[dZ_{2}(t) dZ_{3}(t)] = \rho_{2} dt$ and $ \mathbb{E}[dZ_{1}(t) dZ_{3}(t)] = \rho_{3} dt$. The logarithm of the futures price with time to maturity $\tau$ is given by equation (27) of \cite{schwartz97}:

\begin{equation*}
    \ln{F(t,\tau, S(t), \delta(t), r(t))} = \ln S(t) - \frac{\delta(t)(1-e^{-\kappa \tau})}{\kappa} + \frac{r(t)(1-e^{-
    a \tau})}{a} + C^*(\tau),
\end{equation*}

\noindent where $C^*(\tau)$ is specified in equation (28) of \cite{schwartz97}.

\subsection{Hughen's (2010) stochastic volatility model}

The model of \cite{hughen10}, hereafter \textit{HU-3f}, considers the state vector \({X}(t) = \begin{pmatrix} \ln S(t) & \delta(t) & v(t) \end{pmatrix}^\top \), which, under the risk-neutral measure, satisfies the following SDE:

\begin{align*}
    &d{X}(t) = \left( A + B {X}(t) \right) dt + \Sigma^{1/2}(t,X(t)) d{Z}(t), \\
    &\Sigma(t,X(t)) = \Omega_0 + \Omega_1 v(t),
\end{align*}

\noindent where ${Z}(t)$ is a vector of three uncorrelated Brownian motions, and 

\begin{equation*}
    A = \begin{pmatrix}
        0 \\
        \mu_2  \\
        \mu_3 
    \end{pmatrix}, \quad 
    B = \begin{pmatrix}
        0 & 1 & -1/2  \\
        \kappa_{12} & \kappa_{22} & \kappa_{23}  \\
        0 & 0  & \kappa_{33}
    \end{pmatrix},
\end{equation*} 

\begin{equation*}
    \Omega_0 = 
    \begin{pmatrix}
        0 & s_{12} & \sigma_{13} \vartheta \\
        s_{21} & s_{22} & \sigma_{23} \vartheta \\
        \sigma_{31} \vartheta & \sigma_{32} \vartheta & \sigma_{33} \vartheta 
    \end{pmatrix}, \quad \Omega_1 = \begin{pmatrix}
        1 & \sigma_{12} &  \sigma_{13} \\
        \sigma_{21} & \sigma_{22} & \sigma_{23} \\
        \sigma_{31} & \sigma_{32} & \sigma_{33}
    \end{pmatrix},
\end{equation*}

\noindent where $\Omega_1$ is positive definite and $\Omega_0 - \Omega_1 \vartheta$ is positive semi-definite with $\vartheta \leq 0$. The logarithm of the futures price with time to maturity $\tau$ is given by equation (27) of \cite{hughen10}, that is:

\begin{equation*}
    \ln{F(t,\tau, X(t))} = \alpha(\tau) + \beta(\tau) {X}(t),
\end{equation*}

\noindent where $\alpha(\tau)$ and $\beta(\tau) = \begin{pmatrix}
    \beta_1(\tau) & \beta_2(\tau) & \beta_3(\tau)
\end{pmatrix} $ can be obtained via numerical integration of the following system of ODEs:

\begin{align*}
    \frac{d\alpha(\tau)}{d\tau} = \beta(\tau) A + \frac{1}{2}  \beta(\tau) &\Omega_0 \beta(\tau)^\top, \
    \frac{d\beta_1(\tau)}{d\tau} = \beta(\tau) B_{(1)}, \ 
    \frac{d\beta_2(\tau)}{d\tau} = \beta(\tau) B_{(2)}, \\
    &\frac{d\beta_3(\tau)}{d\tau} = \beta(\tau) B_{(3)} + \frac{1}{2} \beta(\tau) \Omega_1 \beta(\tau)^\top,
\end{align*}

\noindent where $B_{(i)}$ denotes the $i$-th column of $B$, $i=1,2,3$, and the terminal conditions are $\alpha(0)=0,\beta_1(0)=1,\beta_2(0)=0,\beta_3(0)=0$.

\subsection{Yan's (2002) stochactic volatility with jumps model}

The four-factor model of \cite{yan02}, which we denote by \textit{YAN-4f}, comprises the following set of equations under the risk-neutral measure:

\begin{align*}
    &d \ln S(t) =  \left(r(t) - \delta(t) - \nu \mu_J - \frac{\sigma_x^2}{2} - \frac{1}{2} v(t) \right)dt + \sigma_x dZ_{x}(t) + \sqrt{v(t)} dZ_{v}(t) + J dq, \\
    &d\delta(t) =  \left(\mu_\delta - \kappa_\delta \delta(t) \right)dt + \sigma_\delta dZ_{\delta}(t), \\ 
    &dr(t) = \left(\mu_r - \kappa_r r(t) \right) dt + \sigma_r \sqrt{r(t)} dZ_{r}(t), \\
    &dv(t) = \left(\mu_v - \kappa_v v(t) \right) dt + \sigma_v \sqrt{v(t)} dZ_{v}(t) + J_v dq,
\end{align*}

\noindent where \(q\) is a Poisson process with constant intensity \(\nu\), \(J\) and \(J_v \sim \text{Exp}(\theta)\), \(\theta > 0\) denotes the jump size of the log spot price and the volatility, respectively, with \(\ln{(1+J)} \sim N\left(\ln{(1+\mu_J)} - \frac{\sigma_J^2}{2}, \sigma_J^2 \right)\). The Brownian motion increments are correlated as 
$ \mathbb{E}[dZ_{x}(t) dZ_{\delta}(t)] = \rho_{x\delta} dt$ and $ \mathbb{E}[dZ_{x}(t) dZ_{v}(t)] = \rho_{xv} dt$. The logarithm of the futures price with time to maturity $\tau$ is given by:

\begin{equation*}
    \ln{F(t, \tau, S(t), \delta(t), r(t))} = \ln{S(t)} + \beta_0(\tau) + \beta_\delta(\tau) \delta(t) +
    \beta_r(\tau)r(t),
\end{equation*}

\noindent where the expressions for $\beta_0(\tau)$, $\beta_\delta(\tau)$ and $\beta_r(\tau)$ are given in equation (13) of \cite{yan02}. We note that the futures price does not directly depend on the spot volatility. Moreover, only the convenience yield and the spot volatility are correlated with the return process, whereas all the other correlations are set to zero.

\subsection{ Schöne and Spinler (2017) four-factor model} \label{sec_shone}

The four-factor model of \cite{spinler17}, denoted \textit{SS-4f}, uses as state vector ${X}(t) = \begin{pmatrix}
    \ln S(t) & \theta(t) & q(t) & v(t)
\end{pmatrix}^\top$, where $\theta(t)$ represent the long-term level of $\ln S(t)$ and $q(t)$ is a factor related to the cost of carry. \cite{spinler17} formulate their model, under the risk-neutral measure, as follows:

\begin{align*}
    &d{X}(t) = \left[ A + B {X}(t) \right] dt + \Sigma^{1/2}(t,X(t)) d{Z}(t), \\
    &\Sigma(t,X(t)) = \Omega_0 + \Omega_1 v(t),
\end{align*}

\noindent where ${Z}(t)$ is a vector of four uncorrelated Brownian motions and 

\begin{equation*}
    A = \begin{pmatrix}
        0 \\
        {\mu}_2 \\
        {\kappa}_{33}{\mu}_3\\
        {\kappa}_{44}{\mu}_4
    \end{pmatrix}, \quad 
    B = \begin{pmatrix}
        - {\kappa}_{11} & \kappa_{11} & -1 & \kappa_{14} \\
        0 & 0 & 0 & 0 \\
        0 & 0 & - {\kappa}_{33} & 0 \\
        0 & 0 & 0 & - {\kappa}_{44}
    \end{pmatrix},
\end{equation*} 

\begin{equation*}
    \Omega_0 = 
    \begin{pmatrix}
        0 & s_{12} & s_{13} & \vartheta \sigma_{14} \\
        s_{21} & s_{22} & s_{23} & \vartheta \sigma_{24} \\
        s_{31} & s_{32} & s_{33} & \vartheta \sigma_{34} \\ 
        \vartheta \sigma_{41} & \vartheta \sigma_{42} & \vartheta \sigma_{43} & \vartheta \sigma_{44}
    \end{pmatrix}, \quad \Omega_1 = \begin{pmatrix}
        1 & \sigma_{12} & \sigma_{13} & \sigma_{14} \\
        \sigma_{21} & \sigma_{22} & \sigma_{23} & \sigma_{24} \\
        \sigma_{31} & \sigma_{32} & \sigma_{33} & \sigma_{34} \\
        \sigma_{41} & \sigma_{42} & \sigma_{43} & \sigma_{44}
    \end{pmatrix},
\end{equation*}

\noindent where $\Omega_1$ and $\Omega_0 - \Omega_1 \vartheta$ are positive semi-definite with $\vartheta \leq 0$. The logarithm of the futures price with time to maturity $\tau$ is given by equation (11) of \cite{spinler17}, that is

\begin{equation*}
    \ln{F(t,\tau, X(t))} = \alpha(\tau) + \beta(\tau) {X}(t),
\end{equation*}

\noindent where $\alpha(\tau)$ and  $\beta(\tau) = \begin{pmatrix}
    \beta_1(\tau) & \beta_2(\tau) & \beta_3(\tau) & \beta_4(\tau)
\end{pmatrix}$ can be obtained via numerical integration of the following system of differential equations:
\begin{align*}
    \frac{d\alpha(\tau)}{d\tau} = \beta(\tau) A + \frac{1}{2}  \beta(\tau) \Omega_0 \beta(\tau)^\top,  \
    &\frac{d\beta_1(\tau)}{d\tau} = \beta(\tau) B_{(1)}, \
    \frac{d\beta_2(\tau)}{d\tau} = \beta(\tau) B_{(2)},\
    \frac{d\beta_3(\tau)}{d\tau} = \beta(\tau) B_{(3)}, 
    \\ 
     &\frac{d\beta_4(\tau)}{d\tau} = \beta(\tau) B_{(4)} + \frac{1}{2} \beta(\tau) \Omega_1 \beta(\tau)^\top,
\end{align*}

\noindent where $B_{(i)}$ denotes the $i$-th column of $B$, $i=1,2,3,4$, and the terminal solutions are $\alpha(0)= \beta_2(0)=\beta_3(0)=\beta_4(0)=0$ and $\beta_1(0)=1$.

\section{Estimation methodology} \label{sec_method}

Since the variables included in the state vector are typically unobserved, we utilize the estimation framework developed in \cite{duan99}, which is based on the Kalman filter, as it allows us to extract latent variables from a cross-section of observables and to estimate the parameters of the commodity price models (see \cite{schwartz97} and \cite{spinler17}). 

In subsection \ref{sec_futures_estimation}, we illustrate the traditional Kalman filter equations for estimations that rely solely on observed futures prices. In subsection \ref{sec_bond_estimation}, we derive a state-space system and the related filtering equations, which allow for the joint estimation of futures and bond prices. Lastly, subsection \ref{sec_positivity} derives a discretized transition equation for the volatility process, ensuring its positivity.

\subsection{Futures prices} \label{sec_futures_estimation}

At each time $t$ we observe the prices of $H$ futures with different maturities $\tau_i$, for $i=1,\ldots,H$. To proceed, we stack the corresponding pricing equations, derived from \eqref{eq_pde_sol} by adding a measurement error term, resulting in the following representation:

\begin{equation} 
\label{eq_obseq_futures}
         y(t) = \begin{pmatrix}
           \ln F^{(1)}(t,\tau_1,X(t))\\
           \vdots
           \\
           \ln F^{(H)} (t,\tau_{H},X(t)) 
         \end{pmatrix} = 
        \alpha (\tau)+ \beta (\tau) X(t) +
         \begin{pmatrix}
             \varepsilon^{(1)}(t) \\
             \vdots
             \\
             \varepsilon^{(H)}(t)
         \end{pmatrix},
\end{equation}

\noindent where $\alpha (\tau) = \begin{pmatrix}
    \alpha(\tau_1) & \cdots & \alpha(\tau_H)
\end{pmatrix}^\top$, $\beta (\tau) = \begin{pmatrix}
    \beta_1(\tau_1) & \beta_2(\tau_1) &
    \beta_3(\tau_1) & 
    \beta_4(\tau_1) \\
   \vdots &\vdots &\vdots & \vdots\\
    \beta_1(\tau_H) &
    \beta_2(\tau_H) &
    \beta_3(\tau_H) & 
    \beta_4(\tau_H)
\end{pmatrix}$ and $\varepsilon^{(i)}(t)$ represents the measurement error related to logarithm of futures price $i$ with zero mean and standard deviation $\sigma_{\varepsilon_i}$, $i=1,\ldots,H$. This accounts for the possibility of bid-ask spreads, nonsimultaneity of observations and errors in the data. In a correctly specified model, the errors $\varepsilon(t)$ should be serially and cross-sectionally uncorrelated with zero mean.

To obtain the transition equation for the state-space system we need to derive the expressions for the conditional mean and variance of the (unobserved) state variables over a discrete time interval of length $h$, which we set equal to $1/252$ to represent daily data. We discretize  equation \eqref{eq_statevar_p} over $h$ using Euler discretization and we obtain the following transition equation:

\begin{equation} \label{eq_statetransition}
    {X}(t+h) = 
    \widehat{{A}} h + ({I}_4+ \widehat{{B}} h)  {X}(t) + \sqrt{h} {\Sigma}^{1/2}(t,X(t)) {\eta}(t+h),
\end{equation}

\noindent where ${\eta}(t+h)$ is a normally distributed  $(4 \times 1)$ error vector of zero means and unit variances and $I_4$ denotes the $(4 \times 4)$ identity matrix. Given that $\Sigma(t,X(t))$ is a function of $X(t)$, the transition density of equation \eqref{eq_statetransition} will not be Gaussian, so that we have a quasi-optimal Kalman filter. However, as explained in \cite{duan99} and in \cite{ewald21}, the use of this quasi-optimal filter yields an approximate quasi-likelihood function with which parameter estimation can be efficiently carried out. 

The state-space system of equations \eqref{eq_obseq_futures}-\eqref{eq_statetransition} can be directly used in the Kalman filter recursion, which we briefly recall hereafter. Using the set of parameters $\theta_{F} = \{ A, B, \widehat{A}, \widehat{B}, \Omega_0, \Omega_1, \sigma_{\varepsilon_1}, \ldots, \sigma_{\varepsilon_W} \}$, we compute the one-step ahead prediction and variance of $X(t)$ for a step size $h$, conditional to the information at time $t$:

\begin{equation} \label{eq_euler_statevar}
    {X}(t+h|t) = 
    \widehat{{A}} h + \left[{I}_4+ \widehat{{B}} h \right]  {X}(t|t),
\end{equation}

\begin{equation*}
    P(t+h|t) = \left[{I}_4+ \widehat{{B}} h \right] P(t|t) \left[ {I}_4+ \widehat{{B}} h \right]^\top + \Sigma(t, X(t)|t),
\end{equation*}

\noindent where

\begin{equation*}
     \Sigma(t, X(t) |t) = \Omega_0 + \Omega_1 v(t|t).
\end{equation*}

To finish the Kalman filter we need the updating equations:

\begin{align} \label{eq_update_state}
    &X(t|t) = X(t|t-h) + P(t|t-h) \beta(\tau)^\top {V(t|t-h)}^{-1} e(t), \\
    \notag
    &P(t|t) = P(t|t-h) -  P(t|t-h) \beta(\tau)^\top {V(t|t-h)}^{-1} \beta(\tau) P(t|t-h).
\end{align}

\noindent where $e(t) = y(t) - y(t|t-h)$ is the forecast error. We compute then the one-period-ahead prediction and variance of the logarithm of futures prices as:

\begin{align} \label{eq_obs1_fut}
    y(t+h|t) &= \alpha(\tau) + \beta (\tau) X(t+h|t), \\ 
    \label{eq_obs1_fut2}
    V(t+h|t) &= \beta (\tau) P(t+h|t) \beta(\tau)^\top + Q,
\end{align}

\noindent where $Q$ is a $(H\times H)$ diagonal matrix with entries $\sigma^2_{\varepsilon_i}$, $i=1,\ldots,H$. The estimated parameter vector $\theta^*_F$ solves

\begin{equation*}
    \theta^*_{F} = \argmax_{\theta} \sum_{t=1}^N \log L(\theta|e(t) , V(t|t-h) ), 
\end{equation*}

\noindent where $N$ represents length of the time series of futures prices and the day-$t$ log-likelihood is equal to

\begin{equation} \label{eq_likelihood}
    \log L ( \theta | e(t), V(t|t-h) ) = - \frac{1}{2} \left( N \log{2\pi} + \left[ \log \det{(V(t|t-h))} + e(t)^\top V^{-1}(t|t-h) e(t) \right] \right).
\end{equation}

\subsection{Futures prices and bond yields} \label{sec_bond_estimation}


So far, we have shown how to estimate the model using only futures prices. Ideally, as explained in \cite{schwartz97}, the parameters $\theta_F$ of the state vector should be estimated simultaneously from a time series cross-sectional data of futures prices and bond yields. In this section, we illustrate an estimation framework based on a state-space representation that jointly models futures prices and bond yields, allowing for effective estimation of the model parameters. 

Let  $R(t, \tau , {X}(t))$ denote the time $t$ continuously compounded yield on a zero-coupon bond of maturity $\tau$ with price $P(t, \tau, X(t))$: 

\begin{equation} \label{eq_bondyield}
    R(t, \tau , X(t)) = -\frac{1}{\tau} \ln P(t, \tau , X(t)).
\end{equation}

Similarly to our approach for futures prices, we assume that yields for different maturities are observed with errors of unknown magnitudes. Using the bond pricing formula in equation \eqref{eq_sol_bond}, the yield to maturity can be written, after the addition of a measurement error to equation \eqref{eq_bondyield}, as:

\begin{equation} \label{eq_bondprice_err}
         R(t, \tau, X(t)) = - \frac{1}{\tau} \left[ \gamma(\tau) + 
    {\zeta}(\tau) {X}(t) \right] + \psi(t),
\end{equation} 

\noindent where $\psi(t)$ is an error term with zero mean and standard deviation $\sigma_\psi$.

Given that at each time $t$ we observe $H$ futures and $K$ bond yields with different maturities, equations \eqref{eq_obseq_futures} and \eqref{eq_bondprice_err} can be stacked to obtain the following representation:


\begin{equation} \label{eq_obseq_yield}
        y(t) =  \begin{pmatrix}
           \ln F^{(1)}(t, \tau_1, X(t))  \\
           \vdots 
           \\
           \ln F^{(H)}(t, \tau_H, X(t) )  \\
           R^{(1)}(t, \tau_1 , X(t) ) \\
           \vdots 
           \\
            R^{(K)}(t, \tau_K , X(t) ) \\
         \end{pmatrix} =
         \begin{pmatrix}
           \alpha(\tau)  \\
           \gamma(\tau) \\
         \end{pmatrix} + 
         \begin{pmatrix}
            {\beta}(\tau) \\
            {\zeta}(\tau) 
         \end{pmatrix} X(t) +
         \begin{pmatrix}
             \varepsilon^{(1)}(t) \\
             \vdots \\
            \varepsilon^{(H)}(t)\\
              \psi^{(1)}(t) \\
             \vdots \\
             \psi^{(K)}(t) \\
         \end{pmatrix},
\end{equation}

\noindent where  $\gamma (\tau) = 
    \begin{pmatrix} 
     -\frac{\gamma(\tau_1)}{\tau_1} & \cdots
    & -\frac{\gamma(\tau_K)}{\tau_K}
\end{pmatrix}^\top$, $\zeta (\tau) = 
    \begin{pmatrix}
     -\frac{\zeta_1(\tau_1)}{\tau_1} & 
     -\frac{\zeta_2(\tau_1)}{\tau_1} & 
     -\frac{\zeta_3(\tau_1)}{\tau_1} & 
     -\frac{\zeta_4(\tau_1)}{\tau_1} \\
     \vdots &\vdots &\vdots & \vdots\\
      -\frac{\zeta_1(\tau_K)}{\tau_K} & 
     -\frac{\zeta_2(\tau_K)}{\tau_K} & 
     -\frac{\zeta_3(\tau_K)}{\tau_K} & 
     -\frac{\zeta_4(\tau_K)}{\tau_K} 
\end{pmatrix}$ and $\psi^{(j)}(t)$ represents the measurement error related to bond yield $j$ at time $t$ with zero mean and standard deviation $\sigma_{\psi_j}$, $j=1,\ldots,K$. Moreover, given that now we also observe bond yields, we need to consider the parameter set $\theta_{FB} = \{ A, B, \widehat{A}, \widehat{B}, \Omega_0, \Omega_1 , \sigma_{\varepsilon_1}, \ldots, \sigma_{\varepsilon_H}, \sigma_{\psi_1}, \ldots, \sigma_{\psi_K}  \}$. Note that $\theta_{FB}$ contains the same parameters as $\theta_{F}$ along with the standard deviations $\sigma_{\psi_j}$, $j=1,\ldots,K$. In addition, we adjust $Q$ of the observation equation \eqref{eq_obs1_fut2} as a $(H+K)\times (H+K)$ diagonal matrix with entries $\sigma^2_{\varepsilon_i}$ and $\sigma^2_{\psi_j}$, $i=1,\ldots,H$, $j=1,\ldots,K$.

\subsection{Positivity of the state volatility process} \label{sec_positivity}

The Euler discretization scheme in equation \eqref{eq_statetransition} does not guarantee the positivity of the CIR process for the variance process. We address this issue by modifying equation \eqref{eq_statetransition} to utilize an alternative discretization approach. This
adjustment ensures the positivity of the variance state process without enforcing restrictive parameter conditions. We follow the numerical approach developed in \cite{kelly23}, which is to apply a Lamperti transformation to the CIR process and to numerically approximate the related process.

Let us consider the equation for the variance process $v_t$ in \eqref{eq_statevar_p}:

\begin{equation} \label{eq_kelly1}
    dv (t) = \widehat{\kappa}_4(\widehat{\mu}_4 - v (t) ) dt + \sqrt{v(t)} {\Omega^{1/2}_1}_{(4)} dZ(t),
\end{equation}

\noindent where ${\Omega^{1/2}_1}_{(4)}$ denotes the fourth-row of the Cholesky decomposition of $\Omega_1$. By using the Lamperti transform $m(t) = \sqrt{v(t)}$, after an application of It\^{o}’s formula we obtain

\begin{equation} \label{eq_kelly2}
    d m (t) = (\nu m(t)^{-1} - \rho m(t) )dt + \sum_{i=1}^4 \gamma_i  dZ^{(i)}(t),
\end{equation}

\noindent where $\nu = \frac{1}{8}(4 \widehat{\kappa}_4 \widehat{\mu}_4 - \sum_{i=1}^4 (2\gamma_i)^2 )$, $\rho = \widehat{\kappa}_4/2$ and $\gamma_i = \frac{1}{2} {\Omega^{1/2}_1}_{(4)}^{(i)}$, ${\Omega^{1/2}_1}_{(4)} ^{(i)}$ denoting the $i$-th entry of ${\Omega^{1/2}_1}_{(4)}$, $i=1,2,3,4$. Following \cite{kelly23}, let us consider the mesh $\{t_0,t_1,\ldots,t_M\}$ consisting of 
$M+1$ equally spaced points with step size $h$ on the interval $[0,T]$, where $t_0 = 0$ and $t_M = T$. Thus, $t_n = n h$  with $n=0,\ldots,M$ with $h=T/M$. We approximate \eqref{eq_kelly2} using the Lie-Trotter splitting method (see Chapter 2 of \cite{hairer06}):

\begin{equation*}
    m(t+h) = e^{- \rho h} \left( \sqrt{m(t)^2 + 2 \nu h} + \sum_{i=1}^4 \gamma_i \Delta Z^{(i)}(t+h) \right),
\end{equation*}

\noindent where $\Delta Z^{(i)}(t+h) = Z^{(i)}(t+h) - Z^{(i)}(t) \sim N(0, h)$. Then, since $v(t+1) = m^2(t+1) $, we compute

\begin{equation*}
    v(t+h) = e^{ - 2 \rho h} \left( \sqrt{v(t) + 2 \nu h} + \sum_{i=1}^4 \gamma_i \Delta Z(t+h)^{(i)} \right)^2,
\end{equation*}

\noindent or, equivalently,

\begin{align*}
    v(t+h) = e^{ - \widehat{\kappa}_4 h} \Bigg( v(t) + 2 \nu h 
    &+ 2 \sqrt{v(t) + 2 \nu h} \left( \sum_{i=1}^4 \gamma_i  \Delta Z^{(i)}(t+h) \right)  \\
    &\quad + \left( \sum_{i=1}^4 \gamma_i  \Delta Z^{(i)}(t+h) \right)^2 \Bigg)^2,
\end{align*}

\noindent which is non-negative for any $\nu > 0$. We finally proceed to derive the first (conditional) moment as in equation (16) of \cite{kelly23}:

\begin{equation} \label{eq_kelly3}
    v(t + h | t) = e^{ -\widehat{\kappa}_4 h} \left( v(t) + h \left( 2 \nu + \sum_{i=1}^4 \gamma^2_i  \right) \right)  = e^{ -\widehat{\kappa}_4 h} \left( v(t) + h \widehat{\kappa}_4 \widehat{\mu}_4 \right),
\end{equation}

\noindent which we use as a state transition equation for the variance process of the \textit{SRV-4f} model. We note that the Kalman filtering procedure for the benchmark models remains unchanged.

\section{Empirical results} \label{sec_empirical}

Having introduced the models, the data and the estimation methodology, we proceed to describe the calibration results to market data, both in-sample and out-of-sample.

\subsection{Data description} \label{sec_data_descrip}

We use daily observations of crude oil futures prices sourced from Thomson Reuters Eikon Datastream. In particular, we consider light crude oil futures prices quoted on the New York Mercantile Exchange (NYMEX), as discussed in \cite{hughen10}. Specifically, the underlying asset of the futures prices is the West Texas Intermediate (WTI) crude oil spot price, quoted in USD per barrel (USD/BBL)\footnote{BBL is an abbreviation for oilfield barrel.}.



\begin{figure}[h!]
    \centering \includegraphics[width=0.67\textwidth]{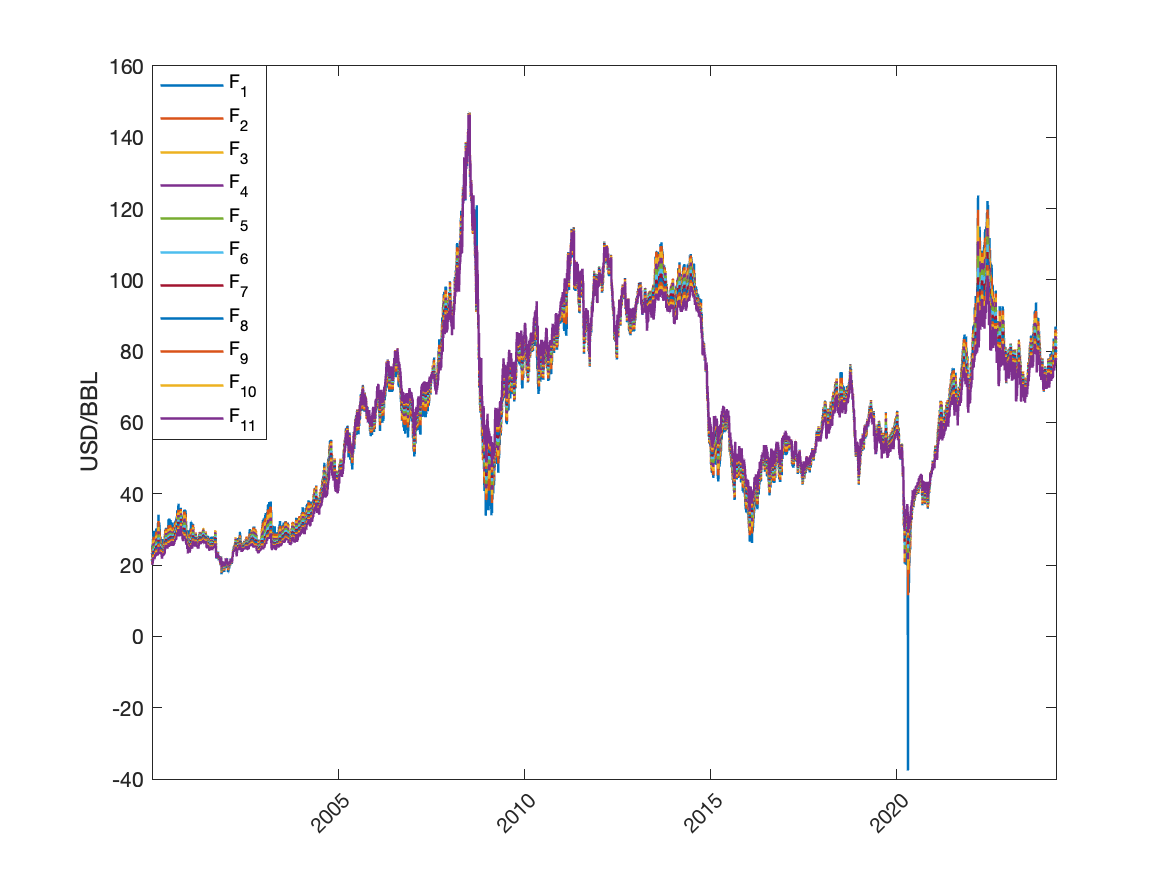}
    \caption{Crude oil futures daily prices (dollars per barrel) for several maturities from January 3, 2000 to April 14, 2024.}
    \label{fig:prezzi_futures}
\end{figure}

Each futures contract is defined by its delivery month, and its trading terminates three business days before the first day of its delivery month. The maturity date for each futures contract is set to the midweek of its delivery month. We select 11 time series of futures prices, organized by their time to maturity: \(F_1\), \(F_2\), \(F_3\), \(F_4\), \(F_5\), \(F_6\), \(F_7\), \(F_8\), \(F_9\), \(F_{10}\), and \(F_{11}\). Here, \(F_n\) represents the contract that is the \(n\)-th month closest to maturity. The sampling period spans 24 years, from January 2000 to April 2024, encompassing a total of 6,333 business days. In Figure \ref{fig:prezzi_futures}, we display the time series of crude oil futures' daily prices for the maturities we considered. We note that on April 20, 2020, both the crude oil spot price and the price of the one-month maturity futures contract
$F_1$ were quoted at negative values. This unprecedented event occurred as the COVID-19 pandemic caused a sharp decline in global petroleum demand, while U.S. crude oil inventories surged. As a result, we opted to exclude $F_1$ from the analysis.

\begin{figure}[h!]
    \centering \includegraphics[width=0.66\textwidth]{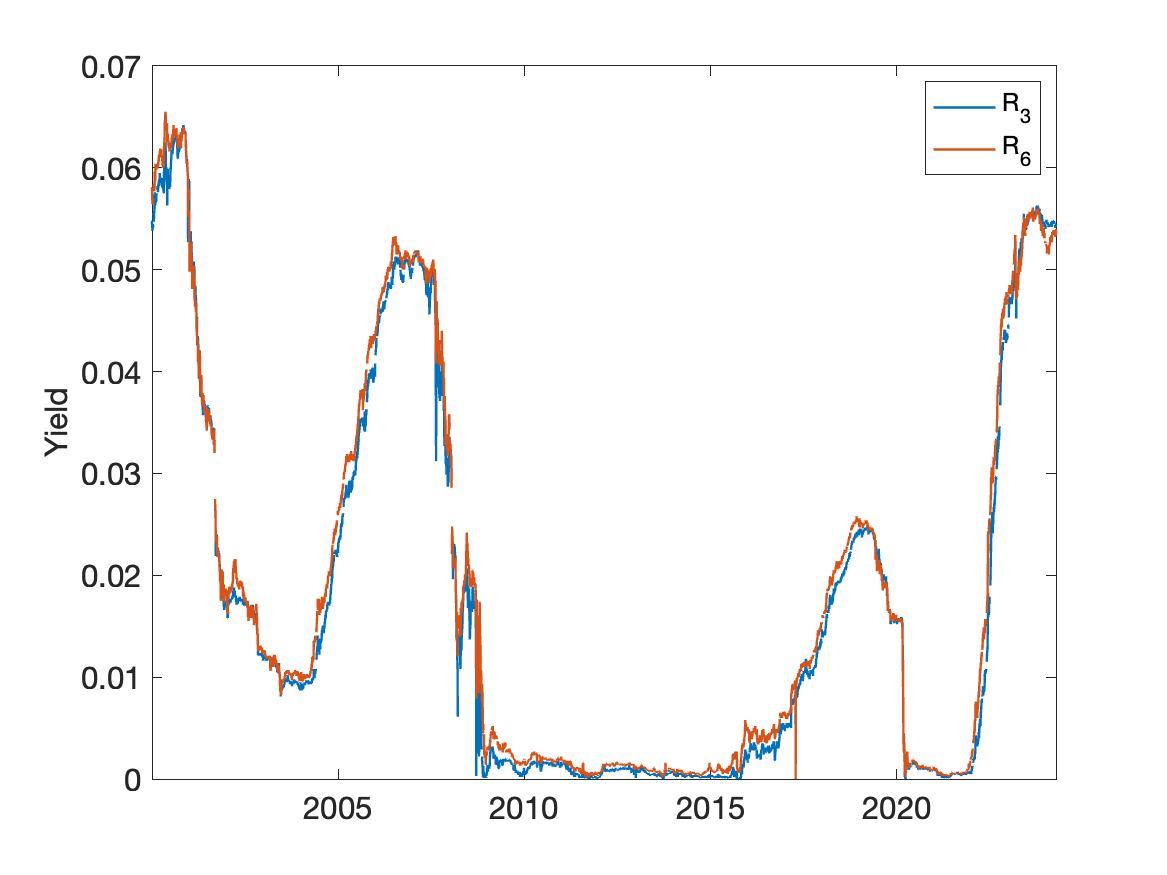}
    \caption{Bond yield daily quotes for two maturities from January 3, 2000 to April 14, 2024.}
    \label{fig:bond_yield}
\end{figure}

Following \cite{schwartz97}, the bond yield data we used consist of U.S. Treasury Par Yield Curve Rates and were gathered from the U.S. Department of the Treasury dataset. We consider the bond yields for two maturities, 3 months and 6 months, denoted as \(R_3\) and \(R_6\), respectively. In Figure \ref{fig:bond_yield}, we display the time series for these bond yields.

\subsection{In-sample analysis}

Table \ref{tab_param_insample} presents the results of the estimation of the \textit{SRV-4f} model. First, we perform an estimation using only futures data, applying equations \eqref{eq_obseq_futures}, \eqref{eq_euler_statevar}, and \eqref{eq_kelly3} with even maturities, specifically \(F_2\), \(F_4\), \(F_6\), \(F_8\), and \(F_{10}\) (so that futures with odd maturities are left for the out-of-sample analysis). As already mentioned, we include bond yield data in the estimation, using equations \eqref{eq_euler_statevar}, \eqref{eq_obseq_yield}, and \eqref{eq_kelly3}. In this joint estimation, bond yields \(R_3\) and \(R_6\) are incorporated alongside the futures prices. In both estimations, the average return on the spot commodity, \(\widehat{\mu}_1\), and the average convenience yield, \(\widehat{\mu}_2\), are positive and significant at standard levels, and, the speed of adjustment coefficients are significant. 


As expected, the correlation between the first two state variables - the spot price and convenience yield - is positive. The correlation between the spot price and volatility is also positive and significant. Thus, several correlations among the state variables are significant, contrary to the assumptions made in \cite{yan02}. Only the correlation between the interest rate and volatility is not significant at any conventional level. Furthermore, it is noteworthy that the significant correlation between the spot price and interest rate change sign when bond yields are included in the estimation. This suggests that the value of a futures contract is sensitive to the interest rate used in its calculation.

\begin{table}[h!]
\centering
\caption{Model estimation results for crude-oil futures.} 
\label{tab_param_insample} 
\begin{tabular}{l S[table-format=2.3(1.3)] S[table-format=2.3(1.3)]}
\toprule
{Parameter} & {\text{$\theta_F$: Only futures}} & {\text{$\theta_{FB}$: Futures and bonds}} \\
\midrule
{$\widehat{\mu}_1$} & 1.670 \,(0.543) & 0.622 \,(0.160) \\
{$\mu_2$} & 4.962 \,(0.631) & -0.151 \,(0.002) \\
{$\widehat{\mu}_2$} & 1.989 \,(0.601) & 0.244 \,(0.107) \\
{$\mu_3$} & 2.027 \,(0.561) & 0.317 \,(4.000e-2) \\
{$\widehat{\mu}_3$} & 2.051 \,(0.527) & -0.053 \,(0.013) \\
{$\mu_4$} & -2.763 \,(0.221) & 1.577 \,(8.000e-3) \\
{$\widehat{\mu}_4$} & 0.227 \,(0.015) & 0.095 \,(0.004) \\
{$\kappa_2$} & 0.222 \,(0.022) & 1.075 \,(0.011) \\
{$\widehat{\kappa}_2$} & -0.723 \,(0.129) & 1.516 \,(0.267) \\
{$\kappa_3$} & 2.712 \,(0.129) & 0.030 \,(4.000e-4) \\
{$\widehat{\kappa}_3$} & 4.711 \,(0.398) & -0.022 \,(0.006) \\
{$\kappa_4$} & 1.557 \,(0.085) & 4.945 \,(0.027) \\
{$\widehat{\kappa}_4$} & 2.075 \,(0.329) & 1.097 \,(0.221) \\
{$s_{12}$} & -0.055 \,(0.008) & 0.045 \,(0.002) \\
{$s_{22}$} & 0.019 \,(0.004) & 0.249 \,(0.038) \\
{$s_{13}$} & -0.171 \,(0.018) & 0.001 \,(1.000e-4) \\
{$s_{23}$} & 0.102 \,(0.018) & 0.001 \,(1.000e-4) \\
{$s_{33}$} & 0.772 \,(0.053) & 0.006 \,(3.000e-4) \\
{$\rho_{12}$} & 0.998 \,(0.056) & 0.572 \,(0.009) \\
{$\rho_{13}$} & 0.444 \,(0.113) & -0.445 \,(0.020) \\
{$\rho_{14}$} & 0.725 \,(0.082) & 0.385 \,(0.017) \\
{$\rho_{23}$} & 0.827 \,(0.124) & -0.933 \,(0.012) \\
{$\rho_{24}$} & 0.583 \,(0.081) & 0.330 \,(0.043) \\
{$\rho_{34}$} & -0.001 \,(2.000e-4) & 0.568 \,(0.317) \\
{$\sigma_{22}$} & 0.375 \,(0.016) & 0.309 \,(0.002) \\
{$\sigma_{33}$} & 0.595 \,(0.146) & 0.015 \,(0.001) \\
{$\sigma_{44}$} & 0.194 \,(0.015) & 0.155 \,(0.007) \\
{$\sigma_{\varepsilon_2}$} & 0.011 \,(0.001) & 0.017 \,(0.001) \\
{$\sigma_{\varepsilon_4}$} & 1.582e-5 \,(6.453e-6) & 4.000e-3 \,(7.145e-5) \\
{$\sigma_{\varepsilon_6}$} & 1.000e-3 \,(1.498e-5) & 6.881e-5 \,(4.807e-6) \\
{$\sigma_{\varepsilon_8}$} & 6.948e-5 \,(1.227e-5) & 2.000e-4 \,(2.530e-6) \\
{$\sigma_{\varepsilon_{10}}$} & 3.539e-5 \,(1.764e-5) & 2.000e-3 \,(6.936e-5) \\
{$\sigma_{\psi_3}$} & {} & 7.535e-6 \,(1.964e-6) \\
{$\sigma_{\psi_6}$} & {} & 2.000e-3 \,(3.467e-5) \\
\bottomrule
\end{tabular}
\vspace{3mm}
\\
\footnotesize
The estimation data includes crude oil futures
\(F_2\), \(F_4\), \(F_6\), \(F_8\), and \(F_{10}\). The standard errors, reported in parentheses, are computed by inverting the negative Hessian matrix evaluated at the optimum parameter values. 
\end{table}

In Table \ref{tab_insample}, we report in-sample results for the different models we considered. The information criteria suggest that the four-factor models,  with the exception of the model in \cite{yan02}, provide a better fit to futures term structure than one, two and three factor models. Moreover, the proposed \textit{SRV-4f} model, when estimated on solely futures data, outperforms the other specifications.

\begin{table}[h!]
\centering
\caption{In-sample model estimation results for crude-oil futures \(F_2\), \(F_4\), \(F_6\), \(F_8\), and \(F_{10}\).}
\label{tab_insample}
\begin{tabular}{S SSS}
\hline\noalign{\smallskip}
     {Model} & {Log.Lik.} & {AIC} & {BIC} \\
     \hline\noalign{\smallskip}
\textit{SCH-1f} & 66.582 & -133.150 & -133.070 \\
\textit{SCH-2f} & 85.473 & -170.920 & -170.810 \\
\textit{SCH-3f} & 92.550 & -185.070 & -184.930 \\
\textit{HU-3f}  & 100.550 & -201.030 & -200.790 \\
\textit{YAN-4f} & 82.786 & -165.530 & -165.350 \\
\textit{SS-4f}  & 109.670 & -192.050 & -191.770 \\
\textit{SRV-4f}  & 135.791 & -271.512 & -271.245 \\
     \hline\noalign{\smallskip}
\end{tabular} 
\end{table}

\begin{figure}
    \centering
    \includegraphics[width=0.7\textwidth]{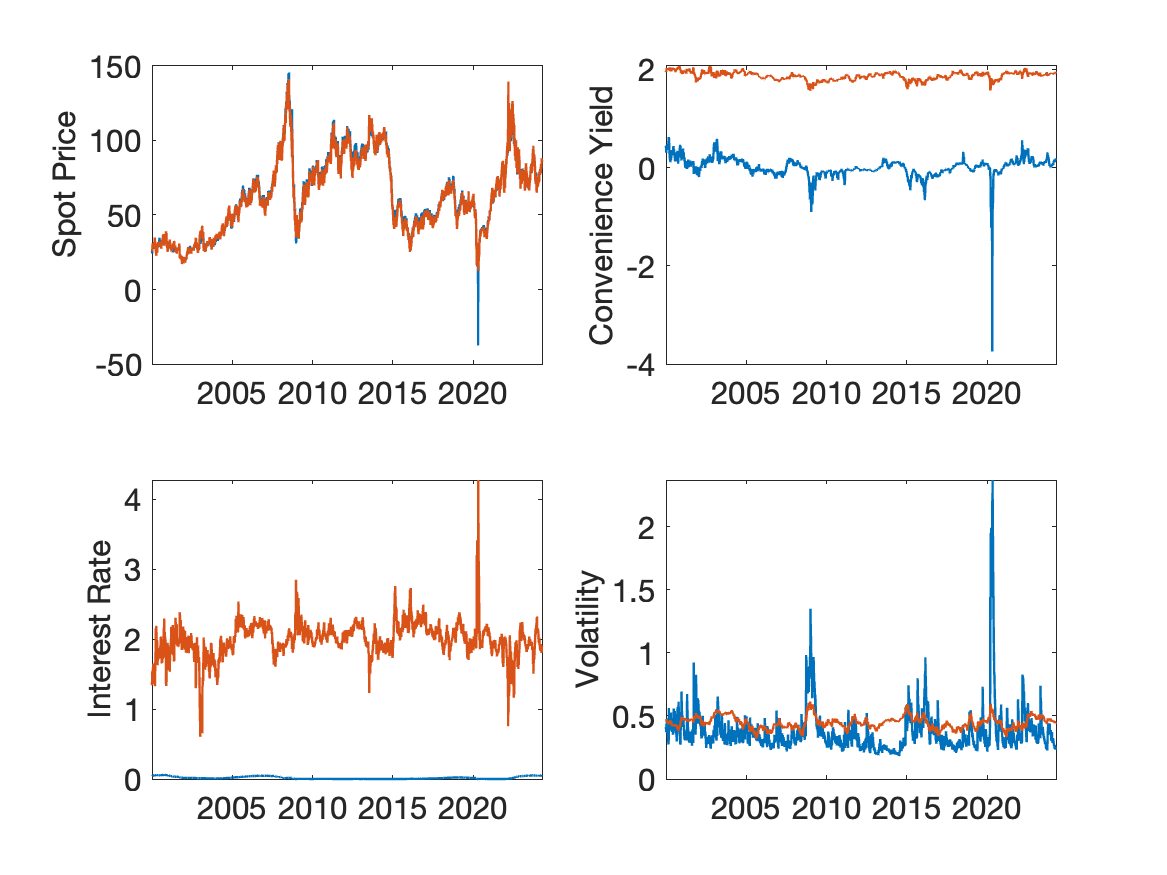}\hfill 
    \caption{Filtered (red) and observable (blue) state variables.}
    \label{fig:state_var_oil_fut}
\end{figure}

Figure \ref{fig:state_var_oil_fut} presents the filtered state variables of the \textit{SRV-4f} model corresponding to the estimation in Table \ref{tab_param_insample}. In this figure, we also include the spot price, which is observed in the market, as well as the other state variables computed using different models or formulas. In particular, the convenience yield has been estimated using the formula in \cite{javaheri03}, while the instantaneous interest rate has been estimated using the Vasicek model. Finally, the volatility of the logarithm of the spot price has been filtered using the (annualized) volatility estimate of the GARCH(1,1) model on log-spot prices, excluding the negative price value from April 20, 2020 (see subsection \ref{sec_data_descrip}).

From the first plot in Figure \ref{fig:state_var_oil_fut}, we observe that the first state variable closely follows the observed crude oil spot price. In the second and third plots, the filtered estimates of the convenience yield and interest rate are not as accurate when compared to the estimates from the models or functions we considered. Finally, the volatility estimate is centered around the mean of the GARCH estimate, but with lower variation.

\begin{figure}
    \centering
    \includegraphics[width=0.7\textwidth]{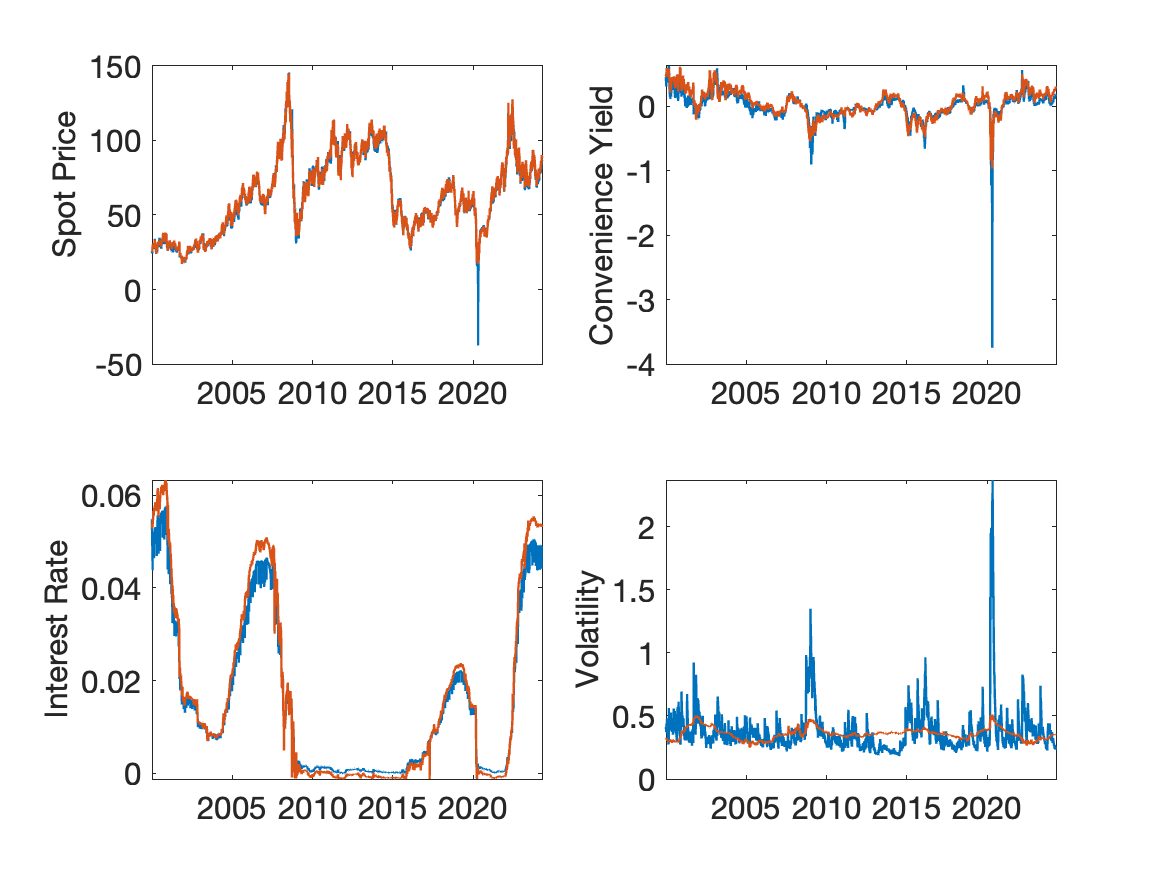}\hfill 
    \caption{Filtered (red) and observable (blue) state variables.}
    \label{fig:state_var_oil_bond}
\end{figure}

Figure \ref{fig:state_var_oil_bond} mirrors Figure \ref{fig:state_var_oil_fut} but it represents results from the joint esimation procedure, using crude oil futures \(F_2\), \(F_4\), \(F_6\), \(F_8\), and \(F_{10}\) in conjunction with bond yields \(R_3\) and \(R_6\). In this case, we can see that the filtered estimate of the interest rate is closer to the Vasicek model estimate. The convenience yield estimate is also more aligned with the one from \cite{javaheri03},  resulting in a more accurate representation of the spot price drift term dynamics.

\begin{figure}[h!]
    \centering
    \includegraphics[width=0.5\textwidth]{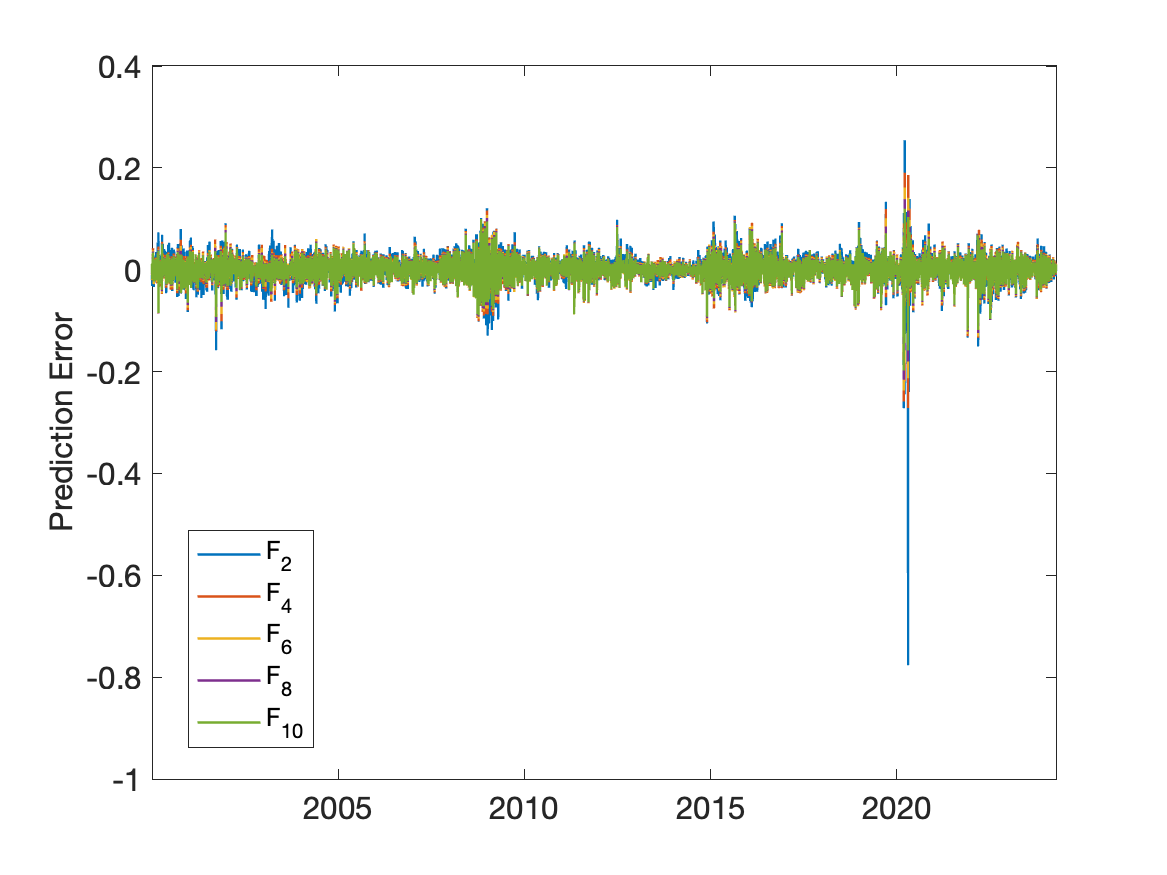}\hfill 
    \includegraphics[width=0.5\textwidth]{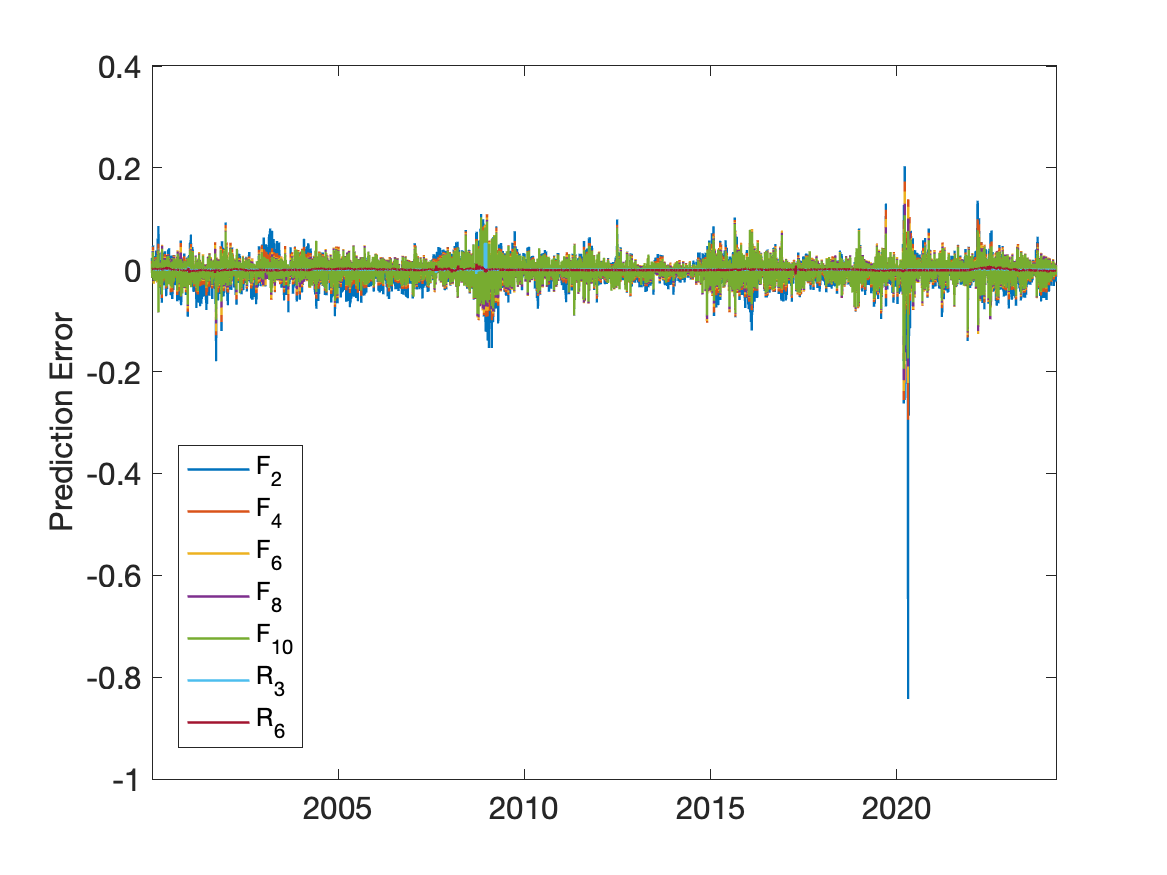}\hfill 
    \caption{ Prediction errors using futures (left panel) and using both futures and bond yields (right panel). The left panel shows errors from the estimation based solely on futures, while the right panel reflects errors from the joint estimation.}
    \label{fig:epsilons}
\end{figure}

In Figure \ref{fig:epsilons}, we display the prediction errors \(e(t)\) for both the estimations based solely on futures and the estimation incorporating futures and bond yields. As we may see, the errors are centered around zero and exhibit volatility clustering, indicating changing levels of uncertainty over time.

\newpage

\subsection{Out-of-sample analysis}

As is standard in the reference literature (see \cite{schwartz97} and \cite{spinler17}), we evaluate futures prices for maturities not included in the estimation. We consider futures prices corresponding to odd maturities, that is, \(F_3\), \(F_5\), \(F_7\), \(F_9\), and \(F_{11}\), and we evaluate them using the parameters from Table \ref{tab_param_insample}.
More precisely, after filtering the state variables and obtaining the coefficients in the futures pricing equation \eqref{eq_pde_sol}, we can price the futures that were excluded from the estimation, that is \(F_3\), \(F_5\), \(F_7\), \(F_9\), and \(F_{11}\). We then compare models using standard metrics, namely root mean square error (RMSE), mean absolute percentage error (MAPE). For a futures contract with maturity $T$, the error metrics are defined as follows:

\begin{align*}
    &\text{RMSE}(T) = \sqrt{ \sum_{t=1}^N \frac{ \left( y(t,T) - \Tilde{y}(t,T,\Tilde{X}(t)) \right)^2 }{N}}, \\
    &\text{MAPE}(T) = \frac{1}{N}\sum_{i=1}^N
    \frac{ |y(t,T) - \Tilde{y}(t,T,\Tilde{X}(t))|}{y(t,T)}, 
\end{align*}

\noindent where $N$ is the length of the time series of futures prices, $y(t,T)$ is the observed log futures price at time $t$ with maturity $T$ and $\Tilde{y}(t,T,\Tilde{X}(t))=\ln{F(t,T-t,\Tilde{X}(t))}$ is the corresponding predicted value, with $\Tilde{X}(t)$ representing the filtered state variable. The results are reported in Table \ref{tab_out_sample}, where we also include the predictive log-likelihood, which is the log-likelihood function evaluated on out-of-sample data using in-sample parameters (for more details, see \cite{gonzales04}). For the \textit{SRV-4f} model, we highlight performance for both the futures-only estimation and the joint estimation involving futures and bonds. In the case of joint estimation, when calculating the predictive likelihood, we used only the futures data, while the model parameters were estimated using both futures and bond data.

The results show the superior performance of the \textit{SRV-4f} model across the different futures maturities, in both the futures-only estimation and the joint estimation with futures and bonds. Specifically, for the futures-only estimation, the RMSE and MAPE futures pricing errors are lower for almost all maturities compared to other models, and the predictive likelihood value is higher. When estimated jointly with bond data, the \textit{SRV-4f} model achieves the lowest average errors for shorter maturities, likely due to the inclusion of short-term bonds in the estimation process.

\begin{table}[h!] 
\centering
\caption{Out-of-sample RMSE and MAPE pricing errors (in percentage) for valuing futures contracts for the considered models, along with the out-of-sample predictive log-likelihood in equation \eqref{eq_likelihood}, computed using the \(F_3, F_5, F_7, F_9, F_{11}\) futures. {\textit{SRV-4f}} (\(\theta_{F}\)) and {\textit{SRV-4f}} (\(\theta_{FB}\)) denote the estimations using only futures and using both futures and bonds, respectively.
}
\label{tab_out_sample}
\resizebox{\textwidth}{!}{
\begin{tabular}{S ccc ccc cc} 
\hline\noalign{\smallskip}
    {Futures} &  {\textit{SCH-1f}} & {\textit{SCH-2f}} & {\textit{SCH-3f}} & {\textit{HU-3f}} & {\textit{YAN-4f}} & {\textit{SS-4f}} & {\textit{SRV-4f}} ($\theta_{F}$) & {\textit{SRV-4f ($\theta_{FB}$)}} \\ \hline\noalign{\smallskip}
{RMSE(3)} & 4.588 & 2.095 & 1.332 & 1.717 & 2.241 & 1.577 & 1.568 & 1.436 \\
{RMSE(5)} & 2.964 & 1.097 & 1.256 & 1.477 & 1.120 & 1.195 & 1.118 & 1.064 \\
{RMSE(7)} & 2.667 & 0.891 & 1.041 & 1.118 & 0.994 & 0.956 & 0.904 & 0.906 \\
{RMSE(9)} & 3.339 & 0.923 & 0.788 & 0.768 & 0.846 & 0.789 & 0.773 & 0.773 \\
{RMSE(11)} & 4.324 & 1.163 & 0.913 & 0.721 & 0.715 & 0.708 & 0.686 & 0.745 \\
\hline\noalign{\smallskip} 
{Mean} & 3.576 & 1.234 & 1.066 & 1.160 & 1.183 & 1.045 & 1.010 & 0.985 \\
     \hline\noalign{\smallskip} \\
         \hline\noalign{\smallskip}
{MAPE(3)} & 0.873 & 0.440 & 0.254 & 0.287 & 0.522 & 0.314 & 0.319 & 0.303 \\
{MAPE(5)} & 0.555 & 0.224 & 0.235 & 0.272 & 0.231 & 0.240 & 0.227 & 0.227 \\
{MAPE(7)} & 0.489 & 0.189 & 0.204 & 0.224 & 0.204 & 0.196 & 0.188 & 0.190 \\
{MAPE(9)} & 0.655 & 0.183 & 0.166 & 0.164 & 0.176 & 0.165 & 0.165 & 0.165 \\
{MAPE(11)} & 0.889 & 0.233 & 0.192 & 0.149 & 0.154 & 0.148 & 0.148 & 0.159 \\
\hline\noalign{\smallskip} 
{Mean} & 0.692 & 0.254 & 0.210 & 0.219 & 0.257 & 0.213 & 0.209 & 0.209 \\
     \hline\noalign{\smallskip} \\
      \hline\noalign{\smallskip} 
     {Predictive Log.Lik.} & 68.631 & 86.761 & 93.987 & 107.420 & 83.555 & 112.160 & 138.862 & 126.680\\
     \hline\noalign{\smallskip}
\end{tabular} 
}
\end{table}

\section{Conclusions} 
\label{sec_conclusions}

We present a novel four-factor model for commodity prices that includes convenience yield, interest rate, and volatility of logarithm of the spot price.  This particular combination has not been explored in existing research, even though these factors play a crucial role in influencing commodity futures prices. Consistent with previous empirical findings, the proposed model allows for time-varying correlations and time-varying risk premiums. We develop a novel estimation framework using the Kalman filter, which enables joint estimation of the term structure of NYMEX crude oil futures prices and U.S. Treasury bond yields. Moreover, we ensure the positivity of the discretized equation for variance process by leveraging an discretization scheme for square-root processes. 

Based on an out-of-sample analysis, our model is statistically preferred over well-known two-factor, three-factor, and even four-factor model specifications that also include stochastic jump factors. Incorporating bond data in the estimation significantly improves the out-of-sample accuracy of short-term futures pricing compared to using futures data alone. Several extensions are possible for our model. One possibility is to add a jump factor to the spot price and examine the effect of the jump risk premium on futures pricing. Another approach is to specify the instantaneous covariance matrix as a function of directly relevant fundamental factors, such as statistical measures of the degree of backwardation. These topics are left for futures research.

\bibliographystyle{unsrtnat}
\bibliography{references}  






\end{document}